# Metro-haul Project Vertical Service Demo: Video Surveillance Real-time Low-latency Object Tracking

**Annika Dochhan[1], Johannes K. Fischer[2], Bodo Lent[3], Achim Autenrieth[4], Behnam Shariati[2], Pablo Wilke Berenguer[2], Jörg-Peter Elbers[4]**

1 ADVA Optical Networking SE, Märzenquelle 1-3, 98617 Meiningen-Dreissigacker, Germany
2 Fraunhofer Institute for Telecommunications Heinrich Hertz Institute, Einsteinufer 37, 10587 Berlin, Germany
3 Qognify GmbH, Werner-von-Siemens-Str. 2 – 6, 76646 Bruchsal, Germany
4 ADVA Optical Networking SE, Fraunhoferstr. 9a, 82152 Martinsried, Germany.
adochhan@advaoptical.com

**Abstract:** We report on the EU H2020 project METRO-HAUL use-case demonstration, including flexible allocation of storage and computing resources in different network locations and deployment of a network slice instance through a programmable multi-layer optical network. © 2020 The Author(s)

## 1. Introduction

Future save and smart cities require intensive increase of video surveillance, which can be supported by future 5G infrastructure. According to [1], the data traffic generated by video surveillance data will increase sevenfold between 2017 and 2022. The EU H2020 5G-PPP project METRO-HAUL [2] aims at architecting and designing cost-effective, energy efficient, agile and programmable metro networks that are scalable for 5G access and future requirements, encompassing the design of all-optical metro nodes, including full compute and storage capabilities, which interface effectively with both 5G access and multi-Tbit/s elastic core networks. The optical layer as well as compute nodes with the required software defined networking (SDN) and network function virtualization (NFV) are developed. The 5G Metro-haul technology developed within the project will be showcased in a future smart city scenario, where security is provided by real-time object recognition and tracking. Simultaneous real-time access to the data of fixed and controllable PTZ (pan-tilt-zoom) cameras allows tracking of objects and persons as well as the automatic recognition of critical events which encompass areas larger than the field of view of a single camera. Dynamically provided network slices with high bandwidth will enable transmission of the video footage of several hundreds of cameras, while the optimum distribution of video analytics computation and storage will reduce the perceived latency. Within this paper we describe the planned demo, which will take place in beginning of 2020 in the Berlin 5G demo testbed in Germany.

## 2. Metro network topology

The scope of the project is the metro area, with typically 20 to 200 km distance between the nodes. Typically, ring or horseshow architectures are employed here, as can be seen in Fig.1. Metro core edge nodes (MCEN) are connected to the core network, while access metro edge nodes (AMEN) are connected in a line to form a horseshoe. At the AMENs, edge data centers (DC) are located, to enable time critical computations close to the end-user, while less latency-sensitive and larger storage requiring functions can be placed in regional or core DC. Various concepts for modulation node architecture were investigated within the project, spanning from low-cost white-box ROADMs [3], to completely filterless architecture [4], using direct detection or coherent transmission [5, 6]. For the demonstration, a semi-filterless architecture is chosen, where add/drop nodes are realized by splitters and couplers with wavelength blockers (WB) in between. Besides enabling high data rates, e.g. 200 Gbit/s net rate at ~30 GBaud with dual polarization (DP) 16QAM, coherent transmission allows the filterless selection and detection of dropped channels by only adjusting the wavelength of the local oscillator laser. Two different types of transponders will be integrated in the demonstration, one is a commercial available CloudConnect QuadFlex[TM] dual-wavelength ADVA transponder ("vendor A" in Fig. 2), capable of 100 Gbit/s to 200 Gbit/s per wavelength, while the other one is a flexible lab device developed by Fraunhofer Heinrich Hertz Institute ("vendor B"), with arbitrary modulation up to DP-64QAM. Vendor A features an OpenConfig [7] interface for software defined control, while vendor B deploys an OpenConfig agent developed within the project. Two different types of WB based ROADMs will be included: One type is based on LCoS technology and EDFAs to cope with the loss of the WB and the transmission fiber, while to the other one is a photonic integrated circuit with semiconductor optical amplifiers (SOAs) to enable loss-less operation [9]. The LCoS-based device will be integrated in ADVA's open line system (OLS), while the other device can be used in a fully disaggregated network with OpenROADM [9] agent. The ADVA OLS offers a north bound Transport API interface (TAPI) [10].

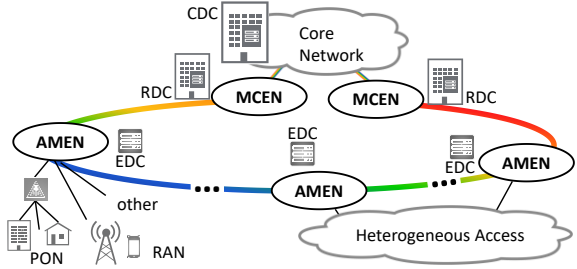

Fig. 1. Metro horseshoe topology. (CDC: central data centre, RDC: regional data centre, EDC: edge data centre, MCEN: metro core edge node, AMEN: access metro edge node).

### 3. Video Surveillance

A smart city video surveillance application requires multiple IP cameras distributed acorss the whole city and connected to several recording servers. One sever might have 100 to 250 cameras connected, either fix cameras, thermal cameras to enable surveillance by night and PTZ cameras to follow objects. In a city-wide installation, the controlling hard and software will be located in a central data center (e.g. RDC or CDC in Fig. 1). In a larger urban area, this central data center will not be in the same network node as the servers, they will be distributed across a city and connected by the optical metro network. Fig. 2 shows the structure of the network with multiple network nodes and cameras. The servers act as core system slaves (CSS) for the video management system and device managers (DM) to control the cameras. They are connected to a core system master (CSM) server, for management and video analytics. A control center client requests video footages from the individual servers and controls cameras manually. Live or archive video could be requested from multiple servers simultaneously, requiring high bandwith through the optical network. For automated and manual control of the cameras, triggered by the analytics or the user, a low latency connection is inevitalble. Analytics can range from simple motion detection to face and even behavior recognition. All these functions of video surveillance require various amounts of bandwidth, storage and computational power.

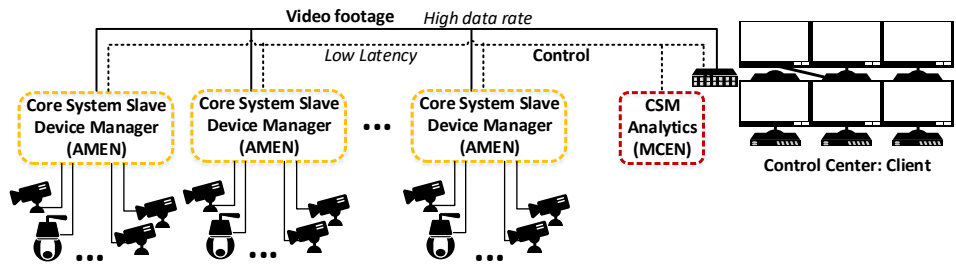

Fig. 2. Video surveillance scenario, multiple cameras are connected to access metro edge nodes (AMENs). Analytics and management is performed at metro edge node (MCEN).

### 4. Metro-haul SDN Architecture

To enable agile and programmable metro networks, service provisioning and traffic dynamicity, a modular and open control, orchestration and management (COM) system using a combination of SDN and NFV technologies that allows the dynamic provisioning of services is implemented [11]. The COM system augments the concept of network control plane with standard interfaces operating across domains to ensure vendor inter-operability. The hierarchical control architecture for this demo is illustrated in Fig. 3. One of the advantages of using a multi-layer approach is the ability to abstract the complexity introduced by the optical technologies inside an optical SDN controller (i.e., ONOS). Furthermore, a parent SDN controller instance can configure and manage the entire metro network with the use of just one uniform south bound interface modeled according to the TAPI information models. Open interfaces export programmability along with unified and systematic information and data modelling. In order to develop a COM system for a multi-layer network, a series of basic services that have to be satisfied: optical layer connection provisioning (i.e., media channel), layer 2 connection service (e.g., VLAN segmentation), IP connectivity between DC nodes (e.g., VXLAN VTEP). Furthermore, a Wide Area Network Infrastructure manager (WIM) should be also be entrusted to provision VNF connectivity between the distributed remote data center installations. The connectivity at the optical layer can be exemplified by creation, deletion, or modification of transparent optical tunnels between two end access points or service interface points (SIPs).

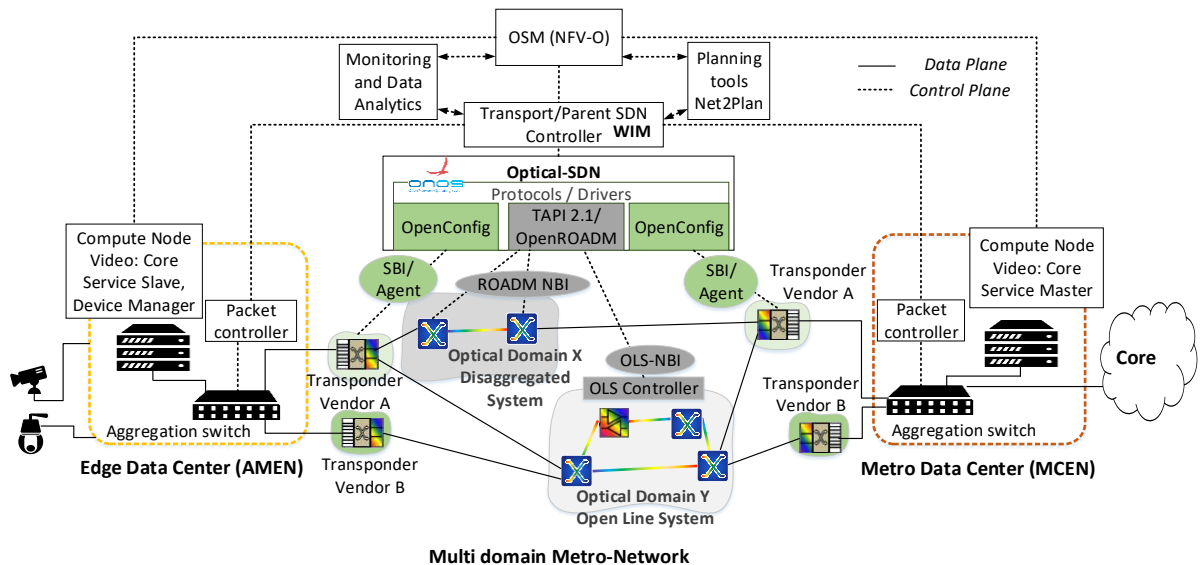

**Multi domain Metro-Network**

Fig. 1: Hierarchical architecture of control, orchestration and management (COM) system for the multi-vendor multi-technology metro network (OSM - Open Source Management and Orchestration, NFV-O – Network Function Virtualization Orchestration, WIM - Wide Area Network Infrastructure manager ,OLS – Open Line System; AMEN – Access Metro Edge Node, MCEN – Metro Core Edge Node; NBI – Northbound Interface; SBI – Southbound Interface; TAPI – Transport API).

All optical transport domains and also the internal optical node structure, are abstracted towards the upper hierarchical control plane level as a single managed topological entity. The proposed architecture also envisions the use of local Virtual Infrastructure Manager (VIM) entities such as OpenStack, deployed in the Data Centers which are capable of instantiating, configuring Virtual Machines (VMs) and attaching virtual interfaces to software switches. The COM system is designed so that the orchestration of such distributed VIMs is accomplished with the use of a central entity in the form of an Open Source MANO (OSM) installation. More exactly, the objective of the OSM is to provide the deployment and service orchestration of the virtual functions for video management and analytics as well as any supporting functions (e.g., video storage database, NAT, firewalling, accounting etc.).The proposed hierarchical and modular software model allows to identify, provision and establish end to end paths between the virtual network functions distributed across multiple administrative and technological domains.

5. **Conclusion**

The paper describes the EU H2020 project METRO-HAUL low-latency object tracking demonstration. The architecture of the hardware and the software is presented, and the video application is explained. Since the demonstration is still in the implementation phase, measurement and demonstration results will be presented later.

6. **Acknowledgement**

This work has received funding from the European Union's Horizon 2020 research and innovation programme under grant agreement No. 761727 (METRO-HAUL).